\documentclass{article}

\usepackage{arxiv}

\usepackage[utf8]{inputenc} % allow utf-8 input
\usepackage[T1]{fontenc}    % use 8-bit T1 fonts
\usepackage{hyperref}       % hyperlinks
\usepackage{url}            % simple URL typesetting
\usepackage{booktabs}       % professional-quality tables
\usepackage{amsfonts}       % blackboard math symbols
\usepackage{nicefrac}       % compact symbols for 1/2, etc.
\usepackage{microtype}      % microtypography
\usepackage{lipsum}
\usepackage{graphicx}
\usepackage{amsmath}
\usepackage{multirow}
\title{\boldmath\bf Electric charge estimation using a SensL SiPM}

\author{
  C.H. Zepeda-Fern\'andez  \\
  Facultad de Ciencias F\'isico Matem\'aticas \\
  Benem\'erita Universidad Aut\'onoma de Puebla \\
  Av. San Claudio y 18 Sur, Ciudad Universitaria 72570, Puebla, Pue.\\
  Cátedra CONACyT, 03940 Ciudad de M\'exico, M\'exico.\\
   \And
    L.F. Rebolledo-Herrera \thanks{corresponding author} \\
    Facultad de Ciencias F\'isico Matem\'aticas\\ 
    Benem\'erita Universidad Aut\'onoma de Puebla \\
    Av. San Claudio y 18 Sur, Ciudad Universitaria 72570, Puebla, Pue.\\
  \texttt{fidel.rebolledo@inaoep.mx} \\
   \AND
   M. Rodr\'iguez-Cahuantzi \\
  Facultad de Ciencias F\'isico Matem\'aticas \\
  Benem\'erita Universidad Aut\'onoma de Puebla \\
  Av. San Claudio y 18 Sur, Ciudad Universitaria 72570, Puebla, Pue.\\
    \AND
   E. Moreno-Barbosa \\
  Facultad de Ciencias F\'isico Matem\'aticas \\
  Benem\'erita Universidad Aut\'onoma de Puebla \\
  Av. San Claudio y 18 Sur, Ciudad Universitaria 72570, Puebla, Pue.\\ 
}

\begin{document}
\maketitle

\begin{abstract}
The silicon photo-multipliers (SiPMs) are commonly used in the construction of radiation detectors such as those used in high energy experiments and its applications, where an excellent time resolution is required for triggering. In most of this cases, the trigger systems electric charge information is discarded due to limitations in data acquisition. In this work we propose a method using a simple radiation detector based on an organic plastic scintillator $2\times2\times0.3$~cm$^3$ size, to estimate the electric charge obtained from the acquisition of the fast output signal of a SensL SiPM model C-60035-4P-EVB. Our results suggest a linear relation between the reconstructed electric charge from the fast output of the SiPM used with respect to the one reconstructed with its standard signal output. Using our electric charge reconstruction method, we compared the sensitivity of two plastic scintillators, BC404 and BC422Q, under the presence of Sr90, Cs137, Co60, and Na22 radiation sources.
\end{abstract}

% keywords can be removed

\section{Introduction}
Silicon Photomultipiers (SiPM) have been widely used during the past two decades in different areas like high energy physics \cite{Simon2019,Garutti2011}, and its medical applications. A clear example is the development of Positron Emission Tomography (PET)\cite{Pizzichemi2019} where the typical photo multiplier tube (PMT) is being replaced by the SiPM technology with the aim to improve its time and spatial resolutions 
\cite{Raczynski2015,Krzemien2015,Ermis2020,Wieczorek2017}.
Since 2013 SensL corporation has developed SiPMs with two signal outputs: Standard and fast \cite{inproceedings}. 
For a $6\times6$~mm$^3$ of SensL C-series SiPM, the standard signal output is characterized by a raise time of 4~ns and a pulse width of 100~ns, while the raise time of the fast signal output is 1~ns with a pulse width of 3.2~ns \cite{SensL2013}.
	
Several works have reported the use of SiPMs \cite{Garutti2011,Pizzichemi2019,Cervi2018,Lamprou2020,Nemallapudi2016}, where the standard output is commonly used to reconstruct the deposited electric charge using the acquired photo current from the anode which is related with the deposited energy in the sensitive material of a radiation detector \cite{Krzemien2015,Lamprou2020,Kuper2017,Lv2018}.  In recent years, the fast signal output has been used to improve the pulse shape discrimination of gammas and fast neutrons \cite{Simon2019, Yu2018}. It has been also shown that exists an equivalent coincidence resolving time (CRT) between the fast and standard output signals of the SensL SiPM \cite{Dolinsky2013}. An application for this fast pulse, is on a detector development with high time resolution as described in \cite{Alvarado2020}.
In this work we use a simple radiation detector based in organic plastic scintillator to study the relation of the reconstructed electric charged using both SensL SiPM output signals.\\
This work is organized as follows. In Section~\ref{2} the methodology of this work is described. In Section~\ref{3} we present the analysis and discussions of the results.  
Finally, in Section~\ref{4} we present our conclusions.

\section{Materials and method} \label{2}
\subsection{Instrumentation}

\begin{figure}[htbp]
    \centering
    \includegraphics[width=0.28\textwidth]{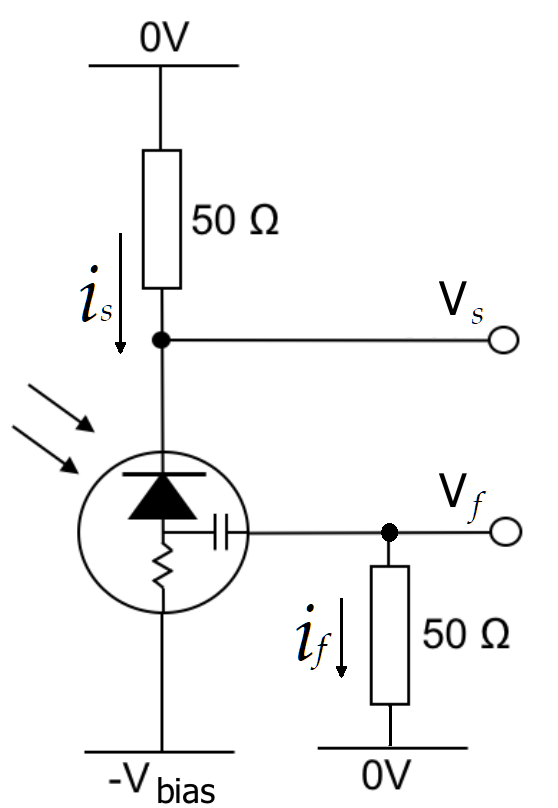}
    \caption{Basic front-end electronics for polarization and acquisition of standard and fast signals.}\label{fig:fe_SiPM}
\end{figure}

A MicroFC-60035 SiPM from SensL with a cell size of 35~$\mu$m, peak wavelength of 420~nm and package size of $6\times6$~mm$^2$ was used. In order to acquire the two signals from this SiPM, we developed a homemade printed circuit board (PCB) of $3 \times 4$~cm, specifically designed for the described SiPM model. The schematic diagram is shown in 
Figure~\ref{fig:fe_SiPM}, where $V_{s}$ and $V_f$ refers to the standard and the fast output signal, respectively.
As described in \cite{SensL2013}, an overvoltage of $V_{br}$+5~V was used to	maximize the photon detection efficiency (PDE) of the SiPM.

We choose BC404 \cite{bc404} and BC422Q \cite{bc422q} plastic scintillator as radiation sensitive materials, with a volume of $20\times20\times3$~mm$^3$. Some scintillation characteristics are shown in Table~\ref{tab:BCchar} \cite{Amcrys}. The BC422Q material was selected with a weight percentage of benzophenone of $0.5\%$.\\

\begin{table}[htbp]
    \caption{Scintillator material properties. \cite{Amcrys}}
    \label{tab:BCchar}
    \centering
    \begin{tabular}{|c|c|c|}
        \hline
        & BC404  & BC422Q  \\
        \hline
        Rise Time (ps) 		& 700 & 110 \\
        Decay Time (ns) 		& 1.8 & 0.7 \\
        Pulse Width, FWHM (ps)	& 2,200 & 360 \\
        \hline 		
    \end{tabular}
\end{table}

\begin{figure}[htbp]
		\centering
		\includegraphics[width=0.4\textwidth]{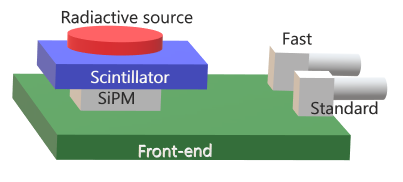}
		\caption{Experimental setup scheme, showing the fast and standard outputs. }\label{SiPM_BC_FE}
\end{figure}

The experimental setup is shown in Figure~\ref{SiPM_BC_FE}, where the SiPM was attached to the center of each plastic scintillator and a radioactive source was located in the opposite face. Four radioactive sources were used: $^{90}$Sr, $^{22}$Na, $^{137}$Cs and $^{60}$Co so, the description for these radioactive sources is shown in Table~\ref{tab:RadSources}.

\begin{table}[htbp]
	\caption{Radiation sources properties.}
	\label{tab:RadSources}
	\centering
		\begin{tabular}{|c|c|c|c|c|}
			\hline
			Source & Radiation  & Gamma 1  & Gamma 2 & Beta \\
			       &  [$\mu$Ci] & [keV]  & [keV] & [keV] \\
			\hline
			$^{60}$Co  & 1.00 & 1173 & 1332 & -\\
			$^{137}$Cs & 0.25 & 662  & - & -\\
			$^{22}$Na  & 1.00 & 511  & 1275 & - \\
			$^{90}$Sr  & 0.10 & - & - & 546 \\
			\hline 		
		\end{tabular}
\end{table}

A Tektronix DPO7054 digital oscilloscope was used for signal acquisition, with a 50~$\Omega$ of coupling impedance and a sampling rate of 10~GS/s. For each radioactive source, $10^{4}$~events were recorded. Each event consists of $2\times10^{4}$~samples to reconstruct the pulse. The reconstruction and the data analysis was made offline with CERN ROOT	software \cite{Brun1997}.\\
	
\subsection{Methods}

\subsubsection{Linear regression}
We reconstruct the electric charge from the fast ($Q_{F}$) and standard ($Q_{S}$) output signals, using the integrals given in equations\eqref{eq:IntSignal} and \eqref{eq:IntSignal}.

	\begin{align}\label{eq:IntSignal}
	Q_{S} &= \int_{t_{i}}^{t_{f}}i_{s}(t)dt = \frac{1}{50}\int_{t_{i}}^{t_{f}}V_{s}(t)dt \\   
	Q_{F} &= \int_{t_{i}'}^{t_{f}'}i_{f}(t)dt = \frac{1}{50}\int_{t_{i}'}^{t_{f}'}V_{f}(t)dt
	\end{align}

If a linear relation between fast and standard signal outputs is assumed, and supposing $g_{x}$ and $g_{y}$ random variables with Gaussian Probability Distribution Functions (PDF), we can introduce the Pearson's correlation coefficient given by \cite{papoullis2002}. 

	\begin{equation} \label{eq:Rxy}
		R_{xy}= \frac{Cov(g_{x},g_{y})}{\sigma_{x}\sigma_{y}}
	\end{equation}
	
where $Cov(g_{x},g_{y})$ is defined as the covariance between random variables $g_{x}$ and $g_{y}$, while $\sigma_{x}$ and $\sigma_{y}$ correspond to the standard deviation of $x$ and $y$ variables, respectively. 
After the described correlation test, a regression line can be adjusted to obtain a linear model based on statistical moments for each stochastic process, as described by the following equation 
	
	\begin{equation} \label{eq:LineRegg}
    y-\bar{y}=\frac{Cov(g_{x},g_{y})}{\sigma_{x}^2}(x-\bar{x})
	\end{equation}
	
It can be rewritten in terms of the correlation coefficient as

	\begin{equation} \label{eq:linRxy}
	y=\frac{\sigma_{y}}{\sigma_{x}}R_{xy}(x-\bar{x})+\bar{y}, 
	\end{equation}
	
which is a standard equation of a straight line  
	
	\begin{equation} \label{eq:recta}
	y=ax + b
	\end{equation}
	
with
	\begin{align}\label{eq:coef}
	&a =\frac{\sigma_{y}}{\sigma_{x}}R_{xy} \\   
	&b =\bar{y}-a\bar{x}      
	\end{align}
	
In this case, $x$ was defined as the charge $Q_{f}$ measured from the standard signal and $y$ as the charge $Q_{s}$ from fast output signal. Therefore, equations \ref{eq:recta} and \ref{eq:coef} can be rearranged in terms of charge,

	\begin{align}\label{eq:linQs}
	&Q_{s}=aQ_{f} + b,\\
	&a =\frac{\sigma_{s}}{\sigma_{f}}R_{fs}, \\   
	&b =\bar{Q_{s}}-a\bar{Q_{f}},      
	\end{align}
	
where $\sigma_{f}$ and $\sigma_{s}$ refer to standard deviation of fast and standard signal charges, respectively. While $R_{sf}$ is the correlation coefficient between $Q_{s}$ and $Q_{f}$.\\
	
As it was mentioned, the linear regression methodology can be applied if the random variables have a Gaussian (PDF) \cite{motulsky2004}, giving the possibility to use the Pearson correlation coefficient. As the PDF acquired from the SensL SiPM are not Gaussian, a non-parametric approach can be used in order to estimate the statistical parameter. This approach makes use of the Spearman correlation index defined by equation \ref{eq:SpearmanRxy} \cite{motulsky2004}
	
\begin{equation} \label{eq:SpearmanRxy}
	R_{rFast,rStd}= \frac{Cov(rFast,rStd)}{\sigma_{rFast}\sigma_{rStd}} 
\end{equation}
	
where $rFast$ and $rStd$ correspond to the ranked fast and standard signals, $\sigma_{rFast}$ and $\sigma_{rStd}$ are the standard deviation of ranked fast and standard signals, viewed as random variables \cite{papoullis2002}. As can be observed in \ref{eq:SpearmanRxy}, the correlation coefficient definition is the same as Pearson correlation, except that the random variables are ranked \cite{papoullis2002}.

\begin{figure}[htb]
	%	\begin{minipage}{16pc}
			\centering
			\includegraphics[width=0.9\textwidth]{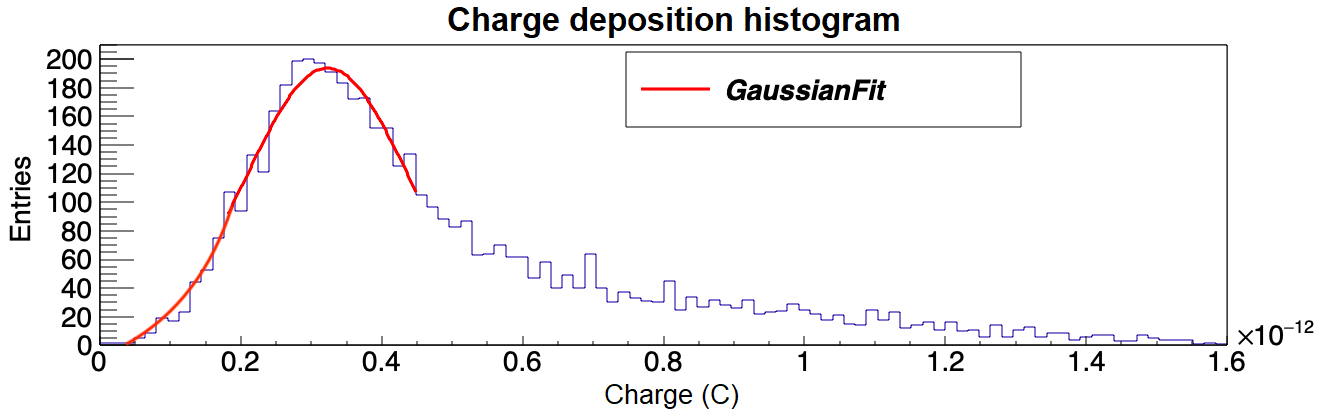}
			\caption{SiPM charge deposition histogram and Gaussian fit for estimation}\label{GaussFitex}
	%	\end{minipage}
	\end{figure}

As described in \cite{looney2011}, this approach does not depend on the used PDFs so, the mean and variance are calculated from a Gaussian distribution, as commonly used.
A second approach to apply the linear regression by fitting a Gaussian PDF to the maximum peaks, observed in the charge deposition histograms from fast and standard signals, as it is shown in Fig~\ref{GaussFitex}; thus, mean and variance are calculated. Moreover, both approaches will be used to demonstrate that standard and fast SiPM signals are highly correlated.

In CERN root software, the integral was performed on windows of $2\times10^3$ samples for fast signal and $1.6\times10^4$ samples for standard signal, because of tail effect.\\

\subsubsection{Two-samples Kolmogorov-Smirnov test} \label{TSKST}		
	It is important to make a quantitative comparison among standard and estimated charge deposition. This can be done using the non parametric Two-Sample Kolmogorov-Smirnov (TSKS) test \cite{hassani2015} which is useful to evaluate whether two underlying one-dimensional probability distributions differ from each other. The Cumulative Density Function (CDF) for each signal is calculated. In the other hand, the test hypothesis can be represented as follows: for a given CDF $F_{s}$, for charge deposition of the standard SiPM output, and a given empirical CDF $F_{est}$, for the estimated charge deposition, the test statistics divergence $D_{n,m}$ can be written as: \cite{hassani2015}   %\ref{eq:TestStatistics} }
	
\begin{equation} \label{eq:TestStatistics}
    D_{n,m} = Max_{x} | F_{s,n}(x)- F_{est,m}(x) |
\end{equation}	

    where $n$ and $m$ correspond to the CDF size of the standard and estimated deposited charge. $Max_{x}$ represents the maximum of the distances set. Moreover, equation \ref{eq:TestStatistics} compares the empirical CDF's from the two charge deposition random variables under test, in order to find out whether both random variables come from same distribution or not. The Kolmogorov-Smirnov (KS) test statistic $\sqrt{n}D_{n,m}$ will help to reject the null hypothesis at level of significance $\alpha$ if $\sqrt{n}D_{n,m}>K_{\alpha}$ for $m,n\to \infty$, where $P(K<K_{\alpha})=1-\alpha$. Thus, the associated $p-value$ is calculated from tables or algorithms, in our case MATLAB was used to apply the KS-test to both CDF's. The resulting $p-value$ is compared with the level of significance $\alpha$ so, the null hypothesis is related to equation \ref{eq:NullHyp} 
\begin{align}\label{eq:NullHyp}
	H_{0}:F_{s}= F_{est} : Failure~to~reject~the~null~hypothesis~at~the~ \alpha-level\\
	H_{1}:F_{s}\not= F_{est} : Rejection~of~the~null~hypothesis~at~the~ \alpha-level
\end{align}    
    	
\section{Analysis and results}\label{3}
	To determine the relation between standard and fast charges, the correlation coefficient was calculated for each radiation source (Co60, Cs137, Na22 and Sr90) and each plastic scintillator (BC404 and BC422Q). As the probability distribution functions for deposited charge from a SiPM are non Gaussian, Pearson correlation index cannot be estimated; therefore, the described approaches from section \ref{2} were used. From the first approach, Spearman correlation coefficient was estimated after data ranking for standard and fast charge estimations \cite{Ma2014}. Then, a Gaussian fit was applied to each distribution for standard and fast charge estimations, resulting on mean and variance calculation and, thus, a Pearson correlation coefficient as reported in 
	Table~\ref{tab:Rxy}.

	\begin{table}[htbp]
	\caption{Correlation $R_{sf}$ between $Q_{s}$ and $Q_{f}$}
	\label{tab:Rxy}
	\centering
		\begin{tabular}{|c|c|c|c|c|}
		\hline
		 - & \multicolumn{2}{c|}{BC404} &\multicolumn{2}{c|}{BC422Q} \\ \hline
		 Source & Pearson & Spearman & Pearson &Spearman \\
		\hline
		Co60  & 0.9686 & 0.9489 & 0.9783& 0.9611\\
		Cs137 & 0.9484 & 0.9157 & 0.9367& 0.9111\\
		Na22 & 0.9549  & 0.9414 & 0.9767& 0.9320 \\
		Sr90 & 0.9658  & 0.9522 & 0.9856& 0.9660 \\
		\hline 		
		\end{tabular}
	\end{table}
	
Correlation coefficient was estimated, resulting on indexes close to one as observed on Table~\ref{tab:Rxy}, which confirms the linear relation hypothesis for fast and standard reconstructed electric charge. Therefore, the described linear regression in Equation~\ref{eq:LineRegg} can be applied to reconstruct the charge distribution for the standard output from the fast signal. This relation is depicted in Figure~\ref{fastslowcharge}, where, the estimated charge correlation for each radioactive source was graphed. {The BC422Q plastic scintillator produces a lower number of photons with respect to BC404. The BC404 has a 68\% of athracene and the BC422Q has a 
19\%~\cite{bc404,bc422q}. Thus, the BC404 emits more photons than BC422Q. A detailed discussion about the properties of plastic scintillators can be found in~\cite{sanjoy}.
	
	\begin{figure}[htbp]
			\centering
			\includegraphics[width=0.49\textwidth]{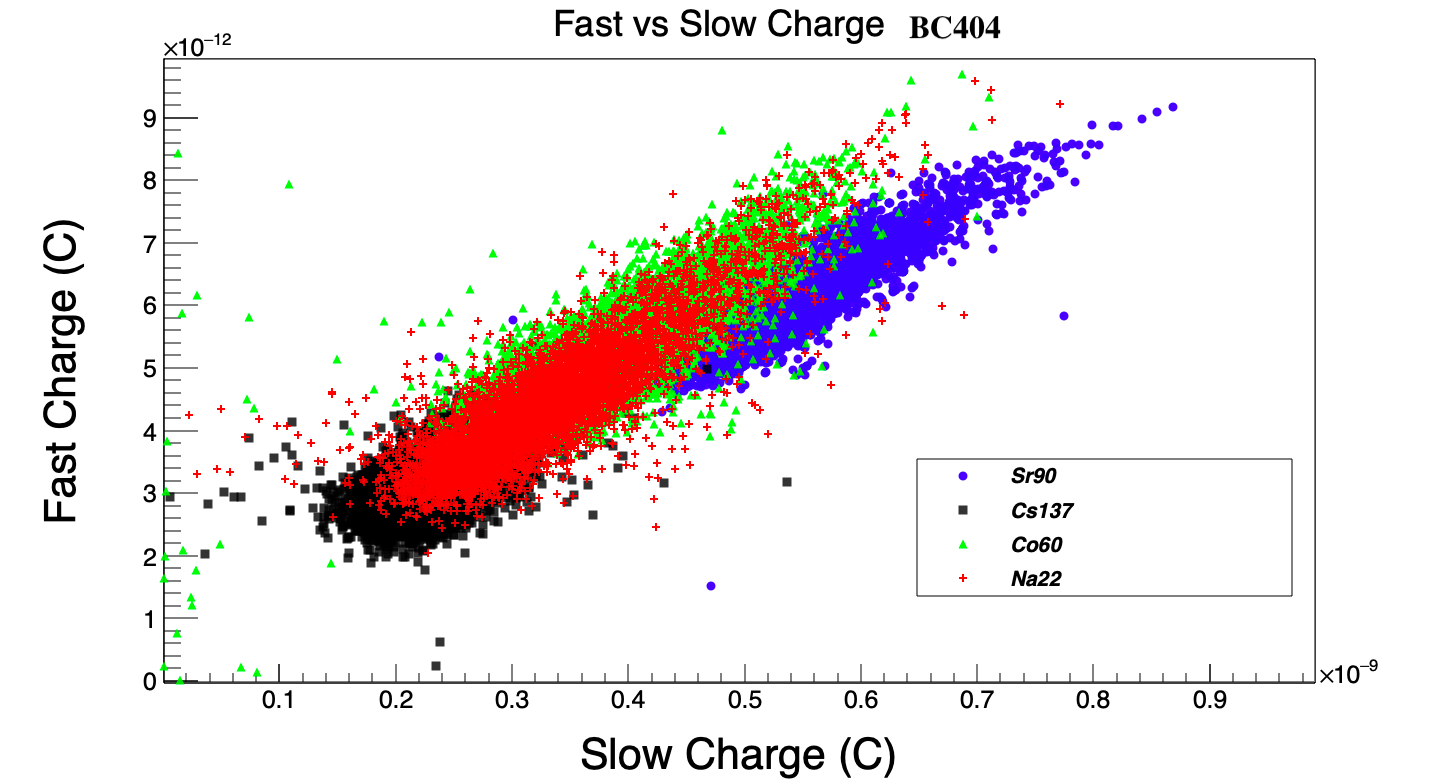}
			\includegraphics[width=0.49\textwidth]{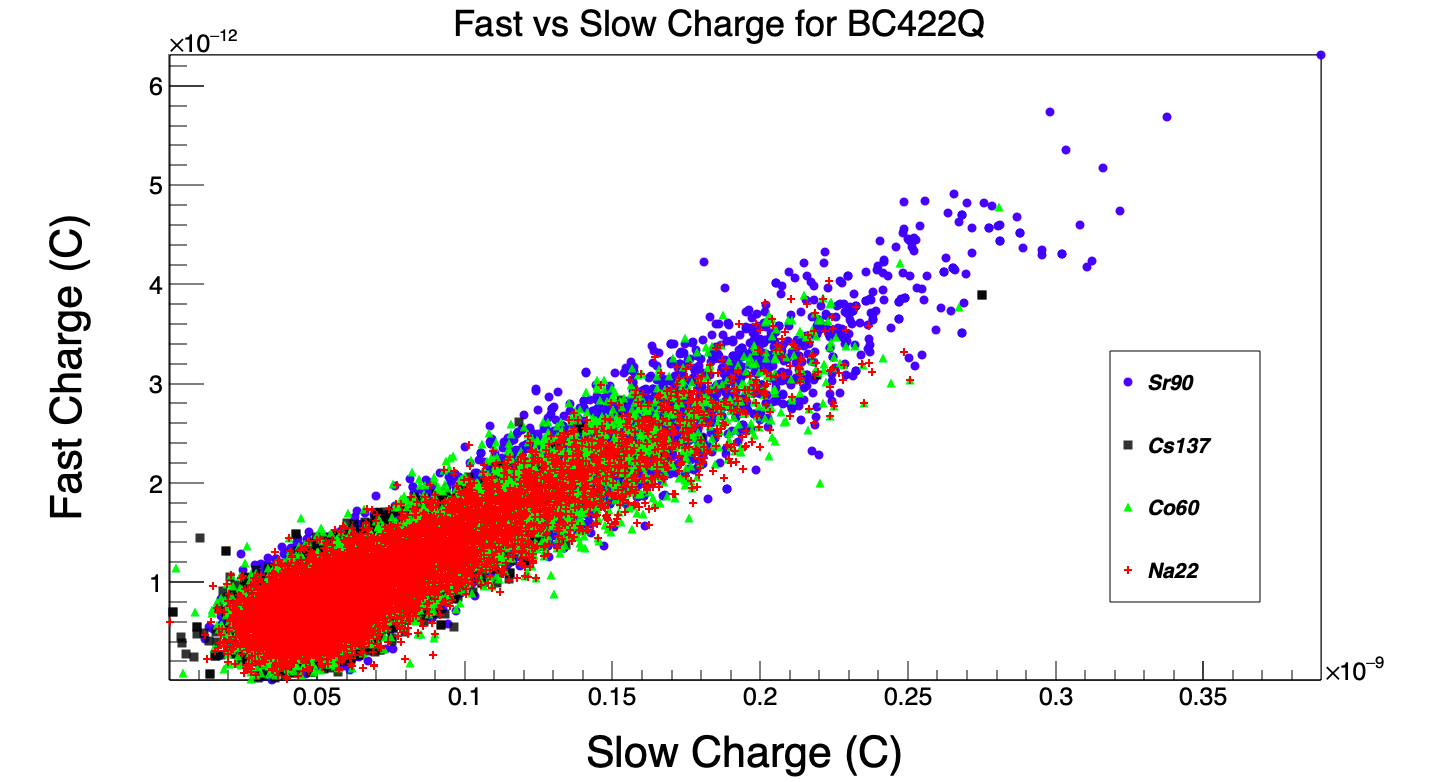}
			\caption{Scatter plot for integrated fast and standard output signals using (left) BC404 and (right) BC422Q scintillators.}\label{fastslowcharge}
	\end{figure}
 	
 	Two orders of magnitude from fast and standard charge can be observed, which is related to the difference of charge deposition among the two variables of interest. Also, it is possible to qualitatively distinguish between the four sources for the case of BC404 scintillator, giving an
 	opportunity for future development of classification algorithm implementation. In particular, $^{90}$Sr and $^{137}$Cs can be clearly separated from $^{22}$Na and $^{60}$Co. For the case of BC422Q scintillator, this source separation seems to be harder to accomplish.

 	The mean and standard deviation are the required statistical momenta for the linear regression model.
 	As an example, in Figure~\ref{fig:fastslow} the charge distribution for the $^{22}$Na radioactive source and both scintillator materials, BC404 and BC422Q, is
 	shown . Doing a Gaussian fit on the the peaks, it is possible to get the required momentum. A particular case is observed for this source, two peaks can be appreciated in fast and standard charge for BC404 scintillator. One of these peaks can be associated to the original gamma from the source and the other can be associated  to a gamma from the pair annihilation from positron emission. For BC422Q scintillator, a single peak is observed for this material. The rest of the sources for both materials have the same shape of one peak.
 	%For the other two sources, the charge distribution has one peak, similar to $^{90}$Sr.

 	\begin{figure}[htbp]
 			\centering
 			\includegraphics[width=0.49\textwidth]{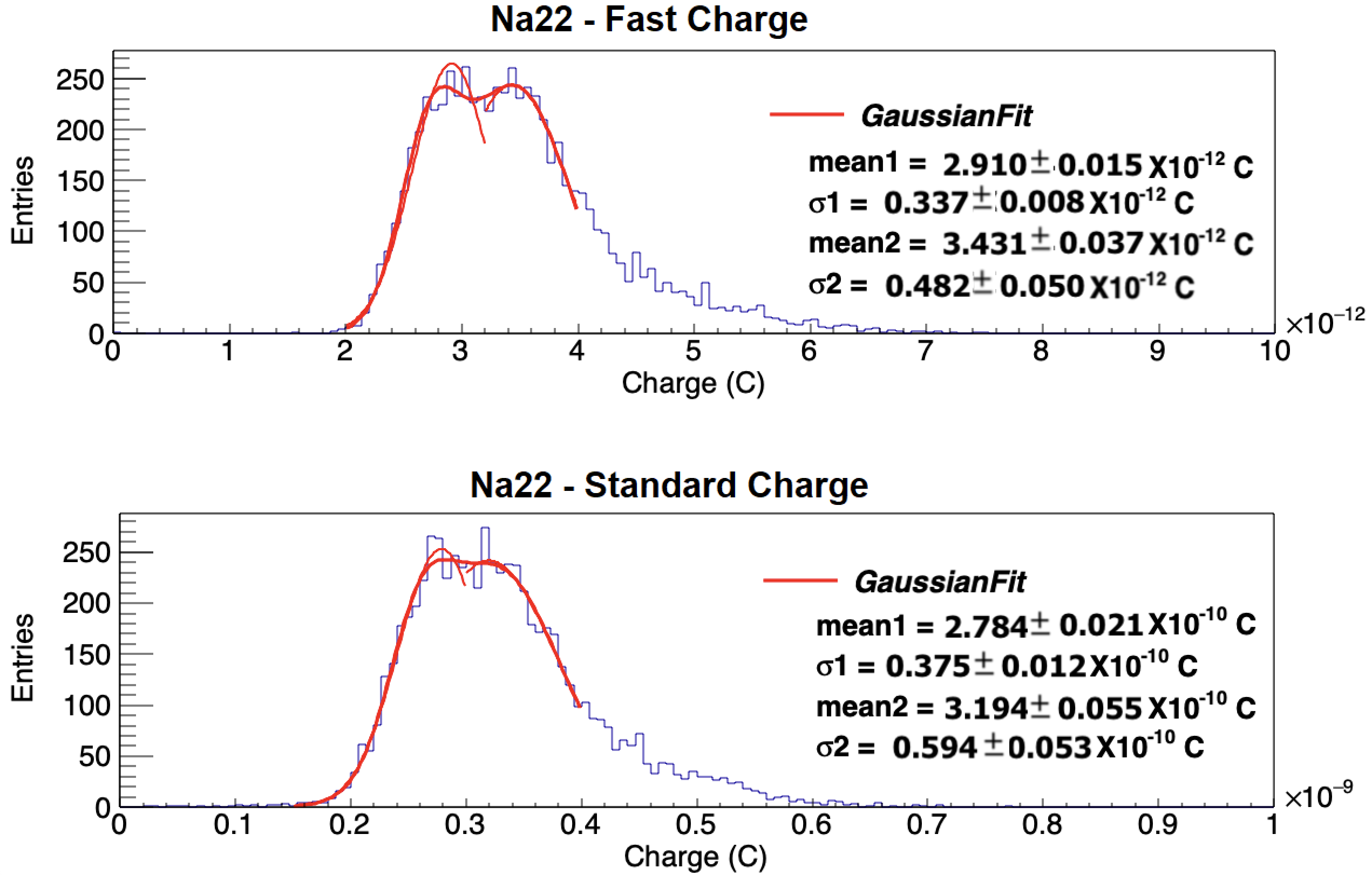}
            \includegraphics[width=0.49\textwidth]{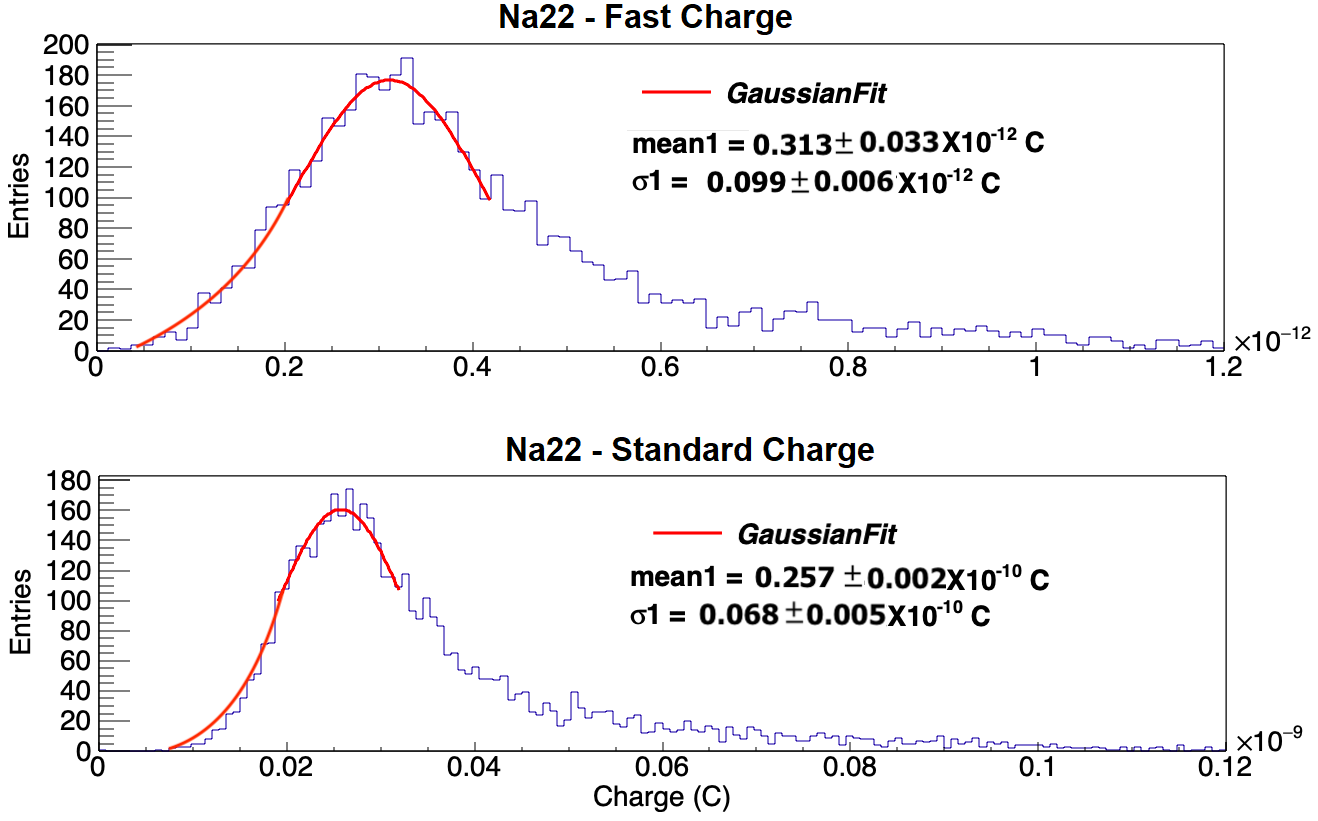}
 		\caption{$^{22}$Na charge deposition comparison and fit for fast and standard outputs for BC404 (left) and BC422Q (right) scintillators.}\label{fig:fastslow}\label{fig:fastslow}
 	\end{figure}
    
    %For BC422Q scintillator, a single peak is observed for $^{22}$Na on the contrary of the reviewed case of BC404 scintillator material. The rest of the sources in both materials have the same shape of one peak.
        
	\begin{figure}[htbp]
	\centering
	\includegraphics[width=0.49\textwidth]{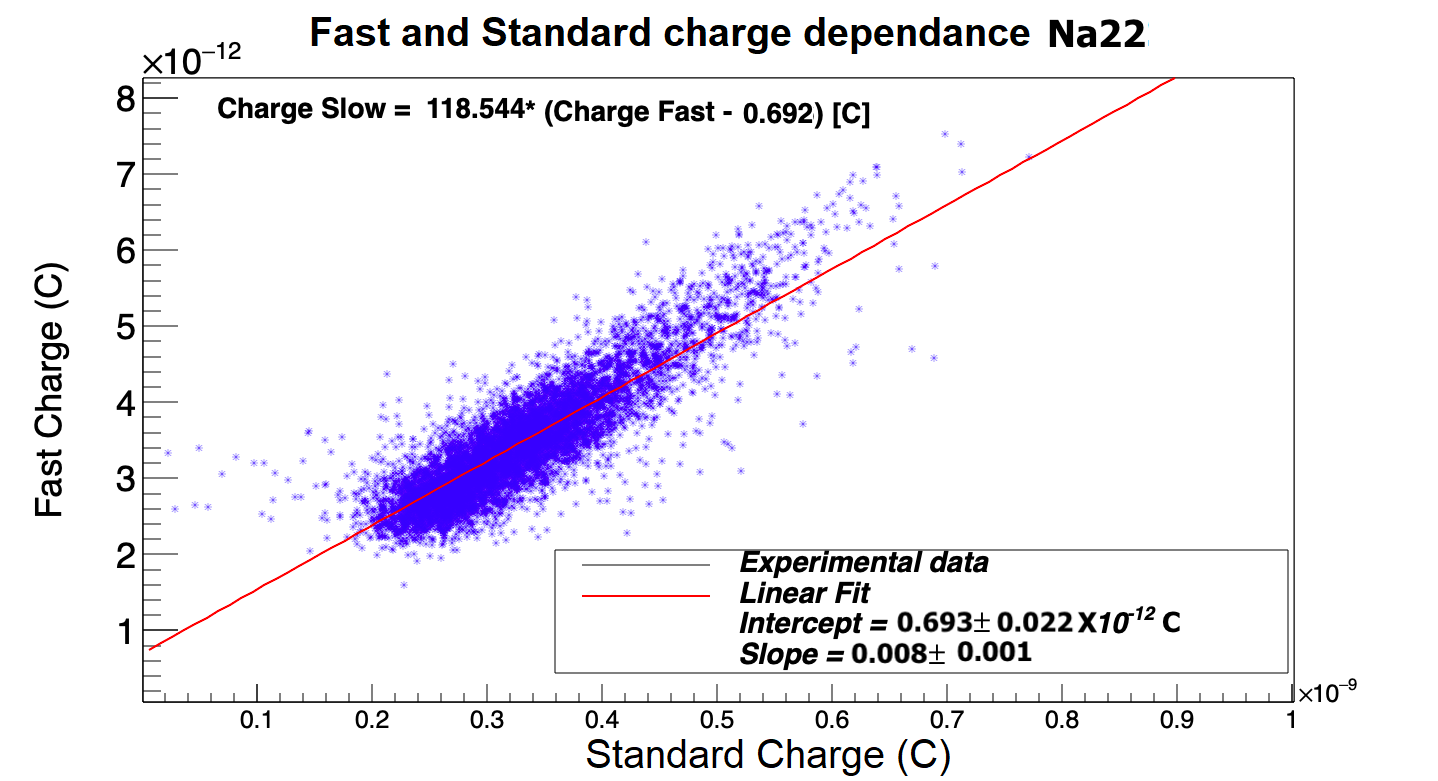}
	\includegraphics[width=0.49\textwidth]{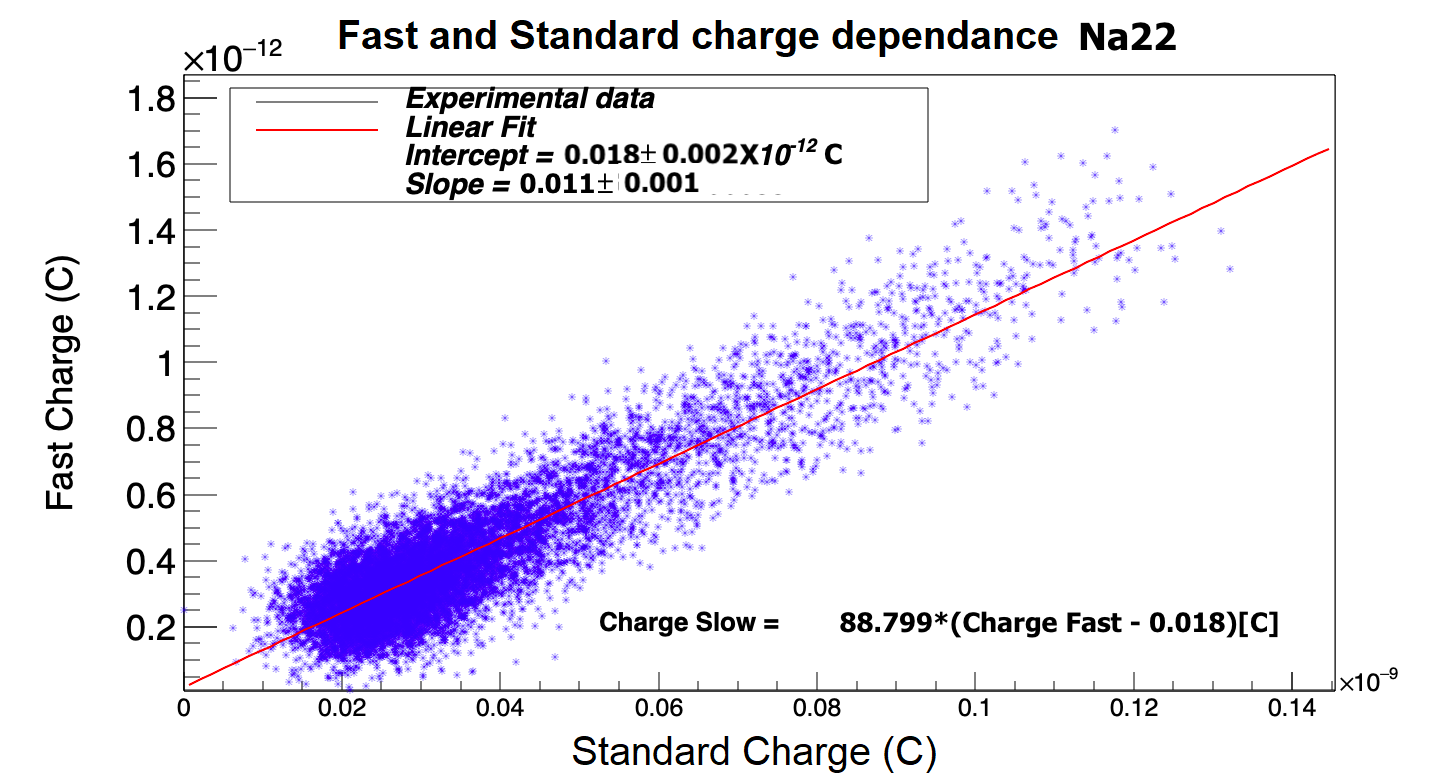}
	%\subfigure[BC404]{}
	%\subfigure[BC422Q]{}
	\caption{Relation between the fast and standard charge for BC404 and BC422Q.}\label{fig:linearfit}
	\end{figure}
	
Once the linear relation between fast and standard charge has been established, all parameters required in the model were calculated. % from acquired data. 
For each random variable, the correlation index $R_{sf}$, the
mean $\bar{Q_{s}}$ and $\bar{Q_{f}}$ and standard deviations 
$\sigma_{s}$ and $\sigma_{f}$ were estimated. 
To exemplify this relation, in Figure~\ref{fig:linearfit} is 
shown  the linear dependence for $^{22}$Na source from BC404 and BC422Q scintillator materials. For a complete reference about the the resulting parameters, the measurements are listed in Tables~\ref{tab:RxyBC404} and \ref{tab:RxyBC422} for BC404 and BC422Q scintillators, respectively. 
Based on results from Tables~\ref{tab:RxyBC404} and \ref{tab:RxyBC422}, the linear regression parameters for each source and material are shown in Tables~\ref{tab:LinReg404} and \ref{tab:LinReg422} for BC404 and BC422Q scintillator, respectively.

	\begin{table}[htbp]
	\caption{BC404 scintillator material linear regression estimated parameters}
	\label{tab:RxyBC404}
	\centering
	\begin{tabular}{|c|c|c|c|c|}
		\hline
		Source& $\bar{Q_{f}}$ &$\sigma_{f}$ & $\bar{Q_{s}}$ & $\sigma_{s}$  \\ 
		- &[$10^{-12}C$] &[$10^{-12}C$] & [$10^{-10}C$]& [$10^{-10}C$]\\
		\hline
		$^{22}$Na$_{peak_1}$&2.910 &0.338 &2.784 &0.374 \\
		$^{22}$Na$_{peak_2}$&3.431 &0.482 &3.194 &0.594 \\
		$^{90}$Sr &5.492 &0.314 &5.127 &0.373 \\
		$^{137}$Cs&2.450 &0.345 &2.297 &0.355 \\
		$^{60}$Co &3.648 &0.259 &3.464 &0.433 \\
		\hline 		
	\end{tabular}
	\end{table}

	\begin{table}[htbp]
	\caption{BC422Q scintillator material linear regression estimated parameters}
	\label{tab:RxyBC422}
	\centering
	\begin{tabular}{|c|c|c|c|c|}
		\hline
		Source& $\bar{Q_{f}}$ &$\sigma_{f}$ & $\bar{Q_{s}}$ & $\sigma_{s}$  \\ 
		- &[$10^{-12}C$] &[$10^{-12}C$] & [$10^{-10}C$]& [$10^{-10}C$]\\
		\hline
		$^{22}$Na &0.311 &0.099 &0.258 &0.068 \\
		$^{90}$Sr &0.323 &0.116 &0.262 &0.074 \\
		$^{137}$Cs&0.333 &0.131 &0.269 &0.082 \\
		$^{60}$Co &0.323 &0.120 &0.263 &0.080 \\
		\hline 		
	\end{tabular}
	\end{table}

	\begin{table}[htbp]
		\caption{Linear regression(BC404).}
		\label{tab:LinReg404}
		\centering
		\begin{tabular}{|c|c|c|c|c|}
			\hline
			Source& a & a (error) & b &b (error) \\ 
			 - & [$10^{-3}$] & [$10^{-5}$] &[$10^{-12}$C] &[$10^{-12}$C]\\
			\hline
			$^{60}$Co  & 7.75 & 6.3 & 1.14 & 0.03\\
			$^{137}$Cs & 6.18 & 9.9 & 1.06 & 0.22\\
			$^{22}$Na  & 8.43 & 9.3 & 0.69 & 0.02\\
			$^{90}$Sr  & 9.06 & 5.5 & 0.95 & 0.03\\
			\hline 		
		\end{tabular}
	\end{table}
	
		\begin{table}[htbp]
		\caption{Linear regression(BC422Q)}
		\label{tab:LinReg422}
		\centering
		\begin{tabular}{|c|c|c|c|c|}
			\hline
			Source& a & a (error) & b &b (error) \\ 
			 - &[$10^{-2}$] & [$10^{-5}$] &[$10^{-12}$C] &[$10^{-15}$C]\\
			\hline
			$^{60}$Co  & 1.16 & 5.3 & 0.014 & 2.30\\
			$^{137}$Cs & 1.01 & 9.4 & 0.052 & 3.10\\
			$^{22}$Na  & 1.13 & 5.5 & 0.017 & 2.22\\
			$^{90}$Sr  & 1.17 & 4.3 & 0.004 & 2.11\\
			\hline 		
		\end{tabular}
	\end{table}
\newpage
Using the respective linear regression from  Tables~\ref{tab:LinReg404} and \ref{tab:LinReg422}, it is possible to reconstruct the original charge distribution from the fast pulse. In Figure~\ref{comp_or_rec} the reconstructed charge distribution for $^{22}$Na and both materials is shown (note that the two peaks for BC404 are reconstructed). We make a statistical analysis to compare the original and reconstructed charge. In Tables~\ref{stat_or_rec_bc404} and \ref{stat_or_rec_bc422q} the
fit parameters for all sources and both materials are shown. \\
As an additional test, we computed the ratio between the integral value of both distributions, reconstructed and original. These values are shown in Table~\ref{ratio_area}.
	
	\begin{figure}[htbp]
  \centering
  \includegraphics[width=0.49\textwidth]{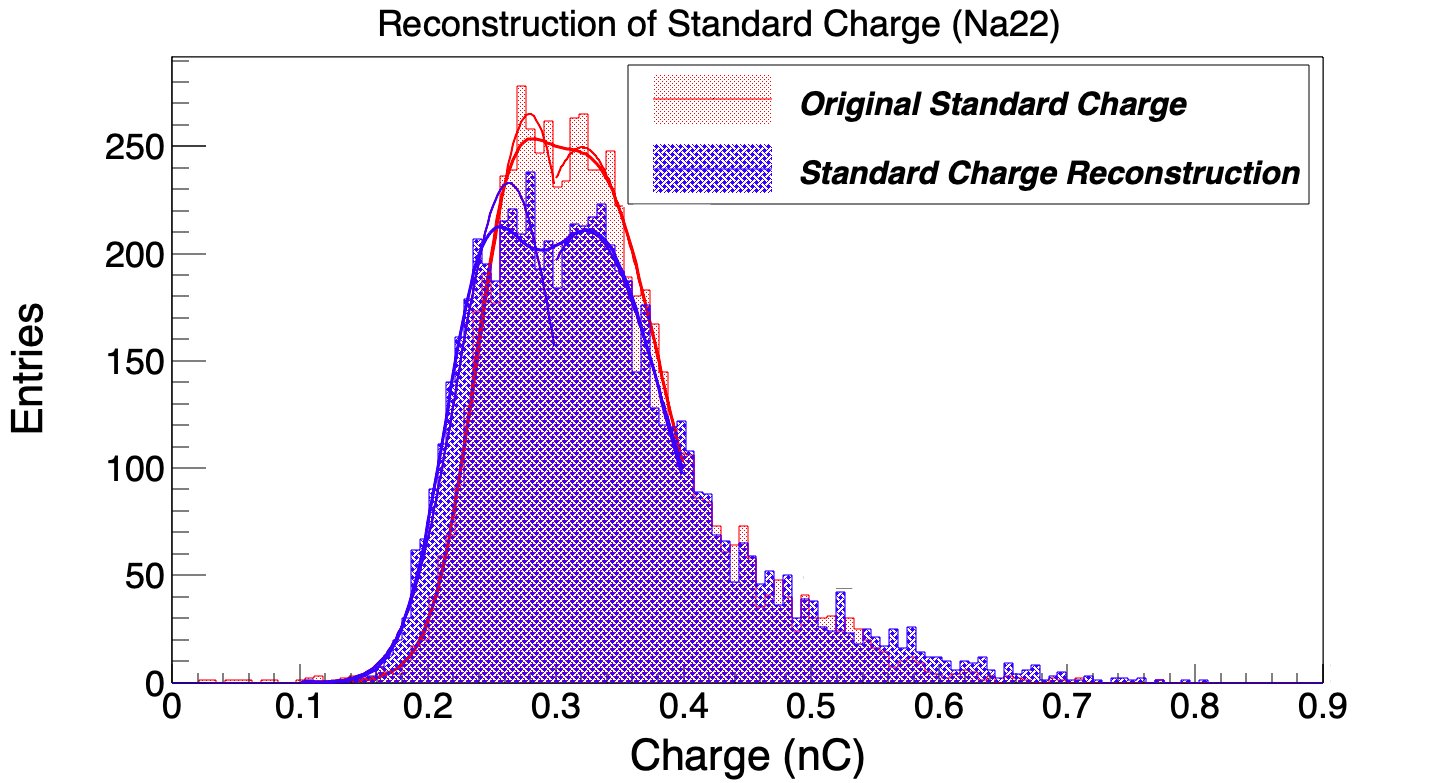}
  \includegraphics[width=0.49\textwidth]{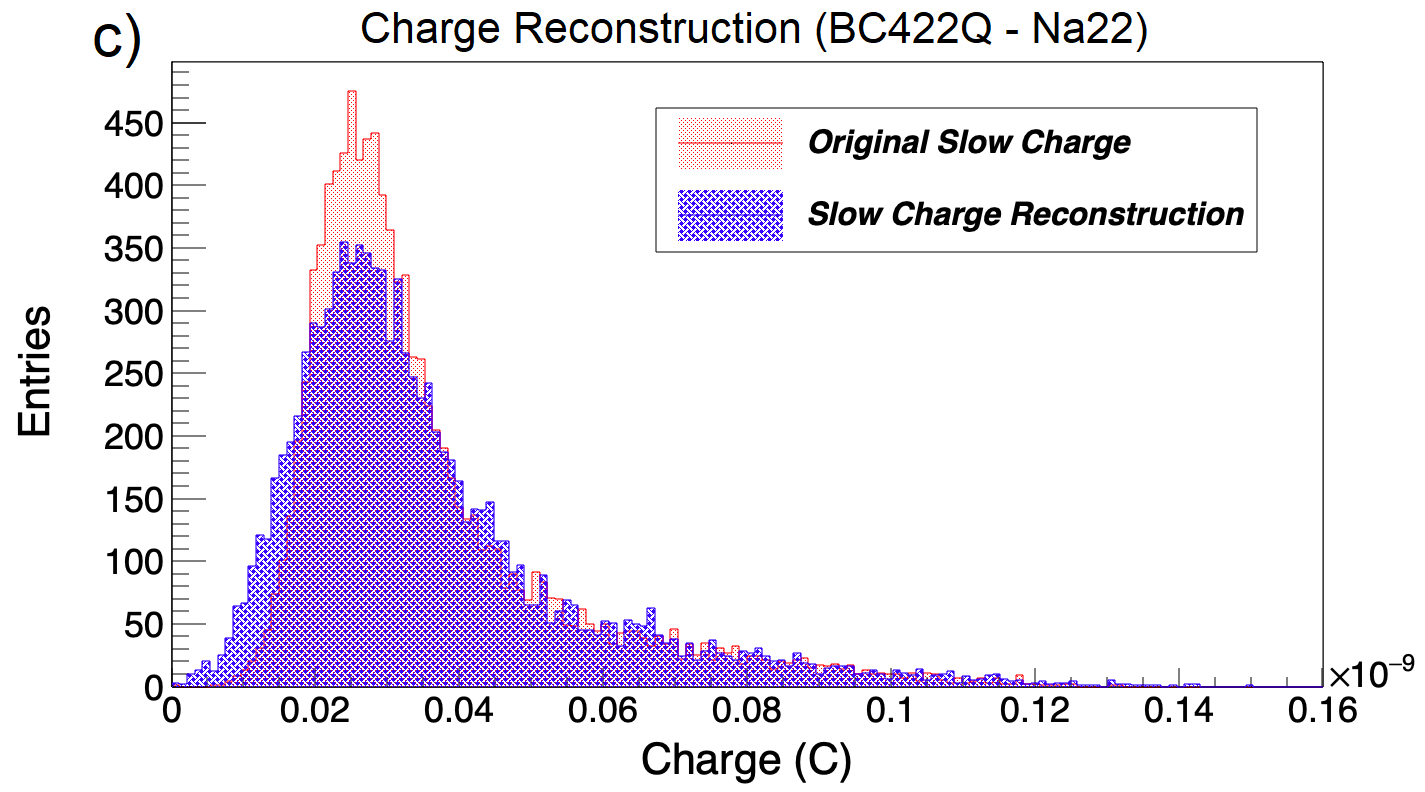}
  \caption{Original and reconstructed charge distributions for Bc404 (left) and BC422Q (right) for $^{22}$Na}\label{comp_or_rec}
\end{figure}

\begin{table}[htbp]
  \begin{center}
    \caption{Gaussian Fit parameters for Bc404.
      $^o$the original charge. $^r$the reconstruction charge.
      $^{1,~2}$the fit parameters for the first and second Gaussian in $^{22}$Na distribution, respectively.}
    \label{stat_or_rec_bc404}
    \begin{tabular}{c|c|c|c|c}
      \textbf{Source} & \textbf{Mean}~(nC) & \textbf{Error Mean}~(nC) & \textbf{$\sigma$}~(nC) & \textbf{Error $\sigma$~(nC)}\\
      \hline
      \multirow{4}{*}{$^{22}$Na} & 0.2795$^{o,1}$ & 0.0023$^{o,1}$ & 0.0380$^{o,1}$ & 0.0012$^{o,1}$\\ 
      &  0.3201$^{o,2}$ & 0.0044$^{o,2}$ & 0.0591$^{o,2}$ & 0.0046$^{o,2}$ \\ % <-- Content of first column omitted.
      &  0.2633$^{r,1}$ & 0.0016$^{r,1}$ & 0.0400$^{r,1}$ & 0.0009$^{r,1}$ \\
      &  0.3231$^{r,2}$ & 0.0053$^{r,2}$ & 0.0622$^{r,2}$ & 0.0063$^{r,2}$ \\
      \hline
      \multirow{2}{*}{$^{90}$Sr}& 0.1276$^o$ & 0.0002$^o$ & 0.0092$^o$ & 0.0001$^o$\\
      & 0.1257$^r$ & 0.0002$^r$  & 0.0088$^r$  & 0.0001$^r$ \\
      \hline
      \multirow{2}{*}{$^{137}$Cs}& 0.2320$^o$ & 0.0004$^o$ & 0.0324$^o$ & 0.0003$^o$\\
      & 0.2257$^r$  & 0.0009$^r$  & 0.0380$^r$  & 0.0007$^r$ \\
      \hline
      \multirow{2}{*}{$^{60}$Co}& 0.3497$^o$ & 0.0008$^o$ & 0.0389$^o$ & 0.0006$^o$ \\
      & 0.3271$^r$  & 0.0009$^r$  & 0.0393$^r$  & 0.0006$^r$ \\
    \end{tabular}
  \end{center}
\end{table}

\begin{table}[htbp]
  \begin{center}
    \caption{Gaussian Fit parameters for Bc422Q.
      $^o$the original charge. $^r$the reconstruction charge.}
    \label{stat_or_rec_bc422q}
    \begin{tabular}{c|c|c|c|c}
      \textbf{Source} & \textbf{Mean~(nC)} & \textbf{Error Mean~ (nC)} & \textbf{$\sigma$~(nC)} & \textbf{Error $\sigma$~ (nC)}\\
      \hline
      \multirow{2}{*}{$^{22}$Na} & 0.0269$^{o}$ & 0.0001$^{o}$ & 0.0066$^{o}$ & 0.0001$^{o}$\\
      &  0.0259$^{o}$ & 0.0002$^{o}$ & 0.0085$^{o}$ & 0.0001$^{o}$ \\
      \hline
      \multirow{2}{*}{$^{90}$Sr}& 0.0273$^o$ & 0.0001$^o$ & 0.0087$^o$ & 0.0001$^o$\\
      & 0.0273$^r$ & 0.0001$^r$  & 0.0086$^r$  & 0.0001$^r$ \\
      \hline
      \multirow{2}{*}{$^{137}$Cs}& 0.0281$^o$ & 0.0001$^o$ & 0.0069$^o$ & 0.0001$^o$\\
      & 0.0266$^r$  & 0.0003$^r$  & 0.0099$^r$  & 0.0002$^r$ \\
      \hline
      \multirow{2}{*}{$^{60}$Co}& 0.0271$^o$ & 0.0001$^o$ & 0.0070$^o$ & 0.0001$^o$ \\
      & 0.0262$^r$  & 0.0002$^r$  & 0.0086$^r$  & 0.0001$^r$ \\
    \end{tabular}
  \end{center}
\end{table}

\begin{table}[htbp]
  \begin{center}
    \caption{Ratio between the area of the original and reconstructed charge distribution.}
    \label{ratio_area}
    \begin{tabular}{c|c|c}
      \textbf{Source} & \textbf{Ratio area distribution for BC404} & \textbf{Ratio area distribution for BC422Q} \\
      \hline
      $^{60}Co$ & 1.0002$\pm$2.2$\times10^{-5}$ & 1.0037 $\pm$2.2$\times10^{-5}$\\
      \hline
      $^{22}Na$ & 1.0001$\pm$1.8$\times10^{-5}$ & 1.0019$\pm$1.1$\times10^{-5}$\\
      \hline
      $^{137}Cs$ & 1.0022$\pm$1.8$\times10^{-5}$ & 1.0028$\pm$1.8$\times10^{-5}$\\
      \hline
      $^{90}Sr$ & 1.0001$\pm$0.1$\times10^{-5}$& 1.0001$\pm$0.1$\times10^{-5}$\\
      \hline
    \end{tabular}
  \end{center}
\end{table}

\newpage
The reconstruction of the charge is performed event by event. Thus, it is possible to make the difference between the original and reconstructed charge value. As an example, in Figure~\ref{Or-Rec}, the plots of such differences for $^{22}$Na for both plastic scintillators are shown. From the fit of these distributions, it is possible to obtain the reconstructed charge resolution, $\sigma$. The resolution values for all the used radiation sources for both materials are shown in Table~\ref{resol_charge}. The best reconstructed charge resolution is obtained for BC422Q plastic scintillator in all the cases.

\begin{figure}[htbp]
\centering
\includegraphics[width=0.49\textwidth]{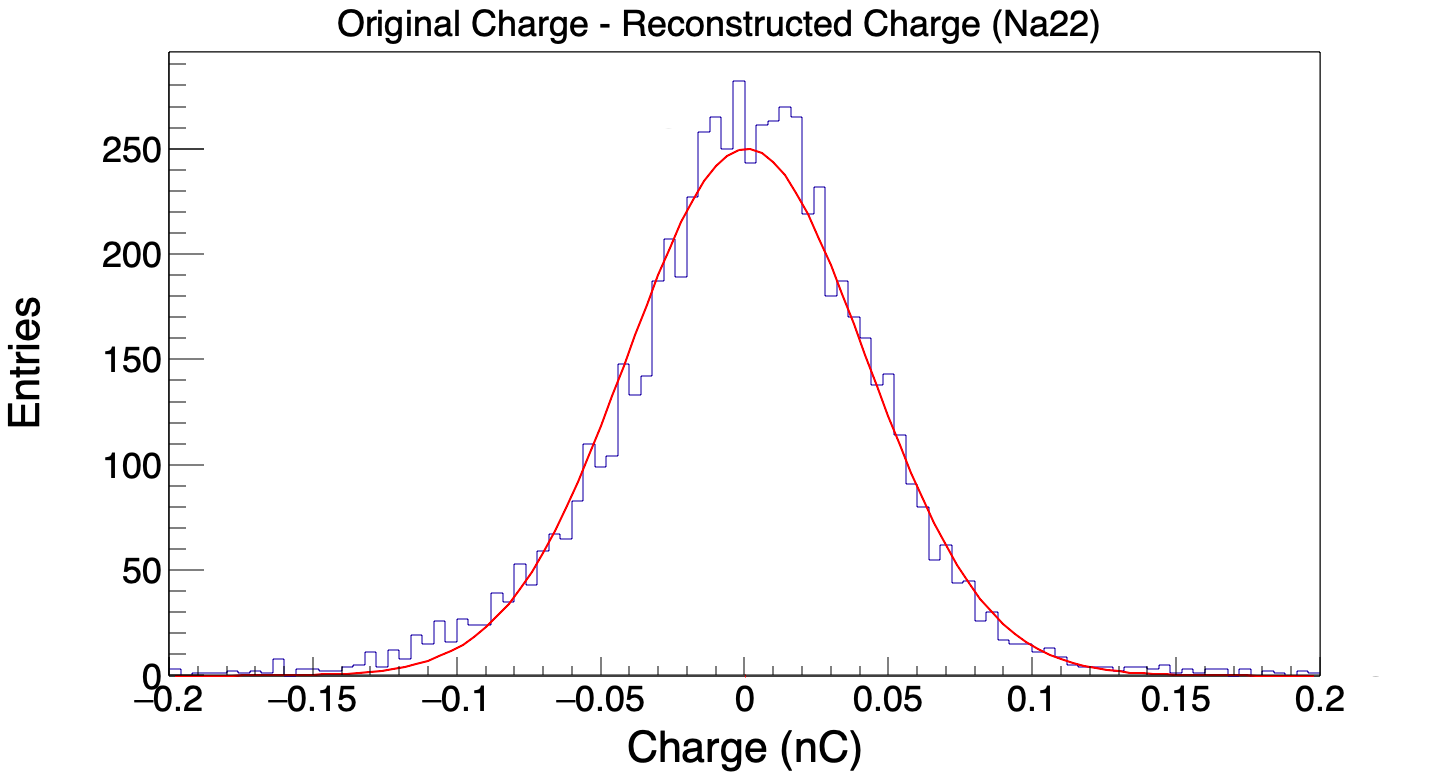}
\includegraphics[width=0.49\textwidth]{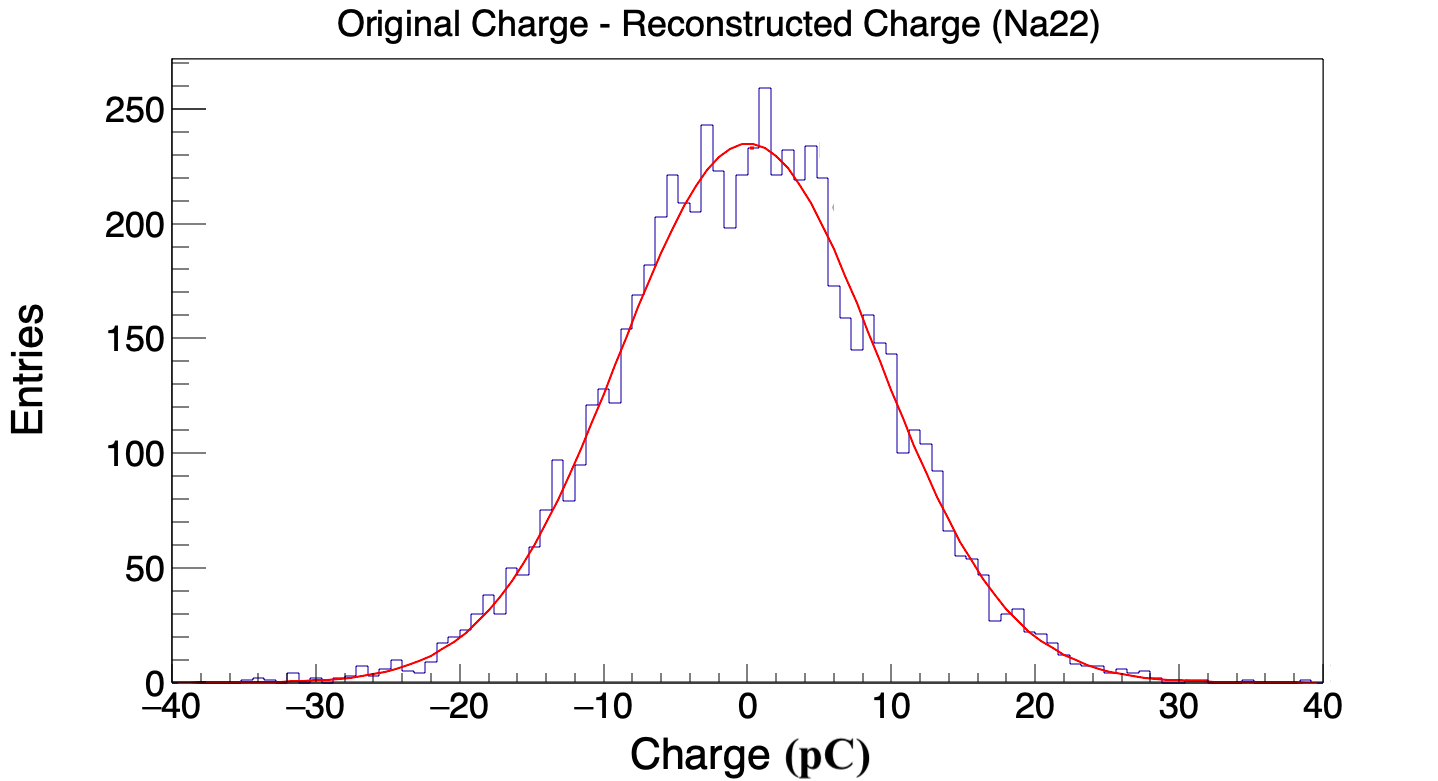}
\caption {Difference distribution between the original charge and the reconstructed charge for $^{22}$Na and  BC404 (left) and BC422Q (right).}
	 \label{Or-Rec}
\end{figure}

\begin{table}[htbp]
  \begin{center}
    \caption{Resolution reconstructed charge for both materials.}
    \label{resol_charge}
    \begin{tabular}{c|c|c}
      \textbf{Source} & \textbf{BC404} & \textbf{BC422Q} \\
      & \textbf{$\sigma$ (pC)} & \textbf{$\sigma$ (pC)}\\
      \hline
      $^{60}Co$ & 47.2$\pm$0.4 & 8.9 $\pm$0.1\\
      \hline
      $^{22}Na$ & 41.4$\pm$0.4 & 8.9$\pm$0.1\\
      \hline
      $^{137}Cs$ & 45.5$\pm$0.4 & 9.3$\pm$0.1\\ 
      \hline
      $^{90}Sr$ & 8.0$\pm$0.1& 8.9$\pm$0.1\\
      \hline
    \end{tabular}
  \end{center}
\end{table}

%The previous analysis to reconstruct the charge deposition from the fast output signal of SiPM and makes a comparison with the information acquired from the standard output signal, in particular, 
We also applied another test to compare the Probability Density Function (PDF) for the standard and  reconstructed charge. The TSKS-test (described in subsection \ref{TSKST}) was applied to both distributions. The CDF's for each radiating source and scintillator material are shown in Figures~\ref{fig:CDFs404} and \ref{fig:CDFs422}. The results of this test are listed in the Table~\ref{tab:TSKStest}. We noted that the corresponding CDF's are equivalent for the standard and reconstructed charge.

	\begin{figure}[htbp]
	\centering
	\includegraphics[width=1\textwidth]{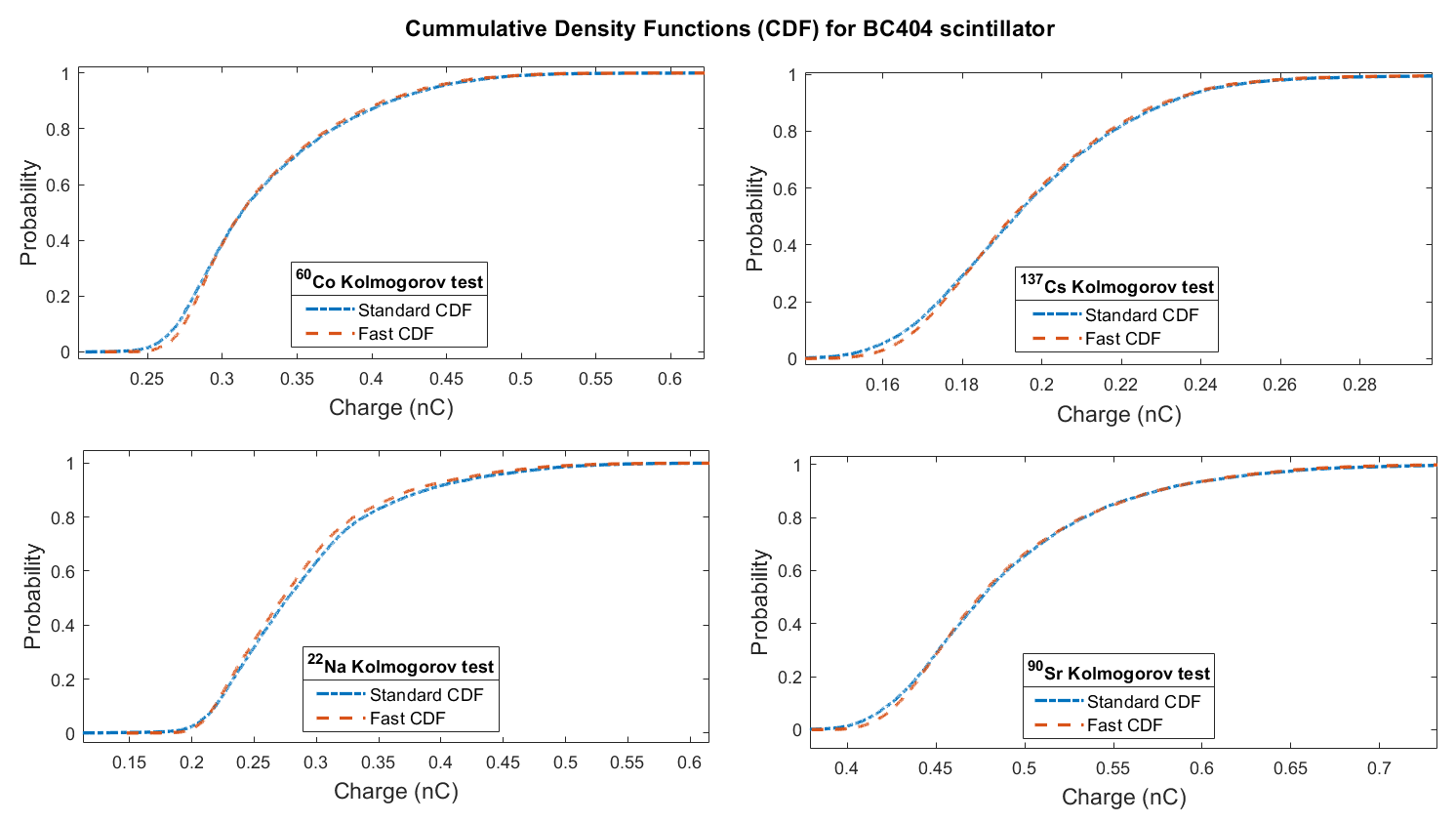}
	\caption{CDF comparative for BC404 scintillator.}\label{fig:CDFs404}
	\end{figure}	
	
	\begin{figure}[htbp]
	\centering
	\includegraphics[width=1\textwidth]{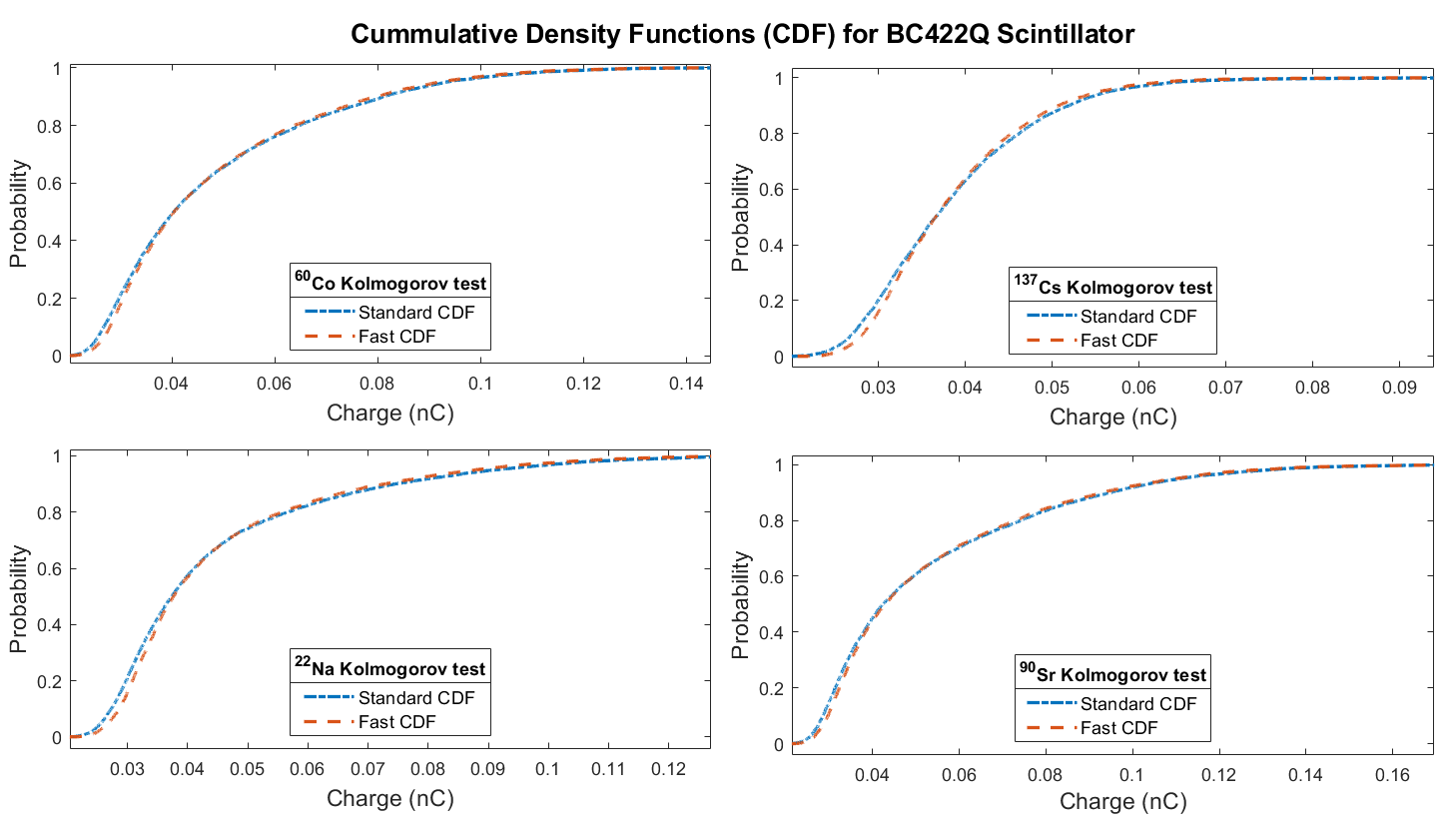}
	\caption{CDF comparative for BC422Q scintillator.}\label{fig:CDFs422}
	\end{figure}	
 
\begin{table}[htbp]
	\caption{Two-samples Kolmogorov test results for BC404 and BC422Q}
	\label{tab:TSKStest}
	\centering
	\begin{tabular}{c|c|c|c|c|c|c}
		\hline\noalign{\smallskip}
			- & \multicolumn{3}{c|}{BC404} &\multicolumn{3}{c}{BC422Q} \\ \hline
		Source & h & p & k & h & p & k\\
		\noalign{\smallskip}\hline
		$^{60}$Co  & 0 & 0.8172 & 0.0090 & 0 & 0.9609 & 0.0076 \\
		$^{137}$Cs & 0 & 0.9975 & 0.0057 & 0 & 0.9983 & 0.0060 \\
		$^{22}$Na  & 0 & 1.0000 & 0.0044 & 0 & 0.9932 & 0.0066 \\
		$^{90}$Sr  & 0 & 1.0000 & 0.0020 & 0 & 0.9901 & 0.0067 \\
		\hline 		
	\end{tabular}
\end{table}
\newpage 
Additionally, the correlation between the mean and $\sigma$ values, obtained from our Gaussian fits for BC404 plastic scintillator material, can be used to distinguish among the different radiation sources used in this work. Accordingly with our results, this is not the case for BC422Q. (see Figure~\ref{meansigma}).
%such correlation is shown. It should be noted that such correlation BC404 plastic scintillator is sensitiveFinally, it was possible to obtain a relation between the mean and $\sigma$ parameters from the Gaussian Fit. This relation is shown in Figure~\ref{meansigma}.  The difference between this two scintillator configurations, is that for BC404, it is possible to make a distinction between the sources. Note that for BC422Q, the Cs137 apparently has the greatest mean, however, by the error values they are consistent. Then, the configuration with the BC422 scintillator does not allow	

	\begin{figure}[htb]
	\centering
	\includegraphics[width=0.49\textwidth]{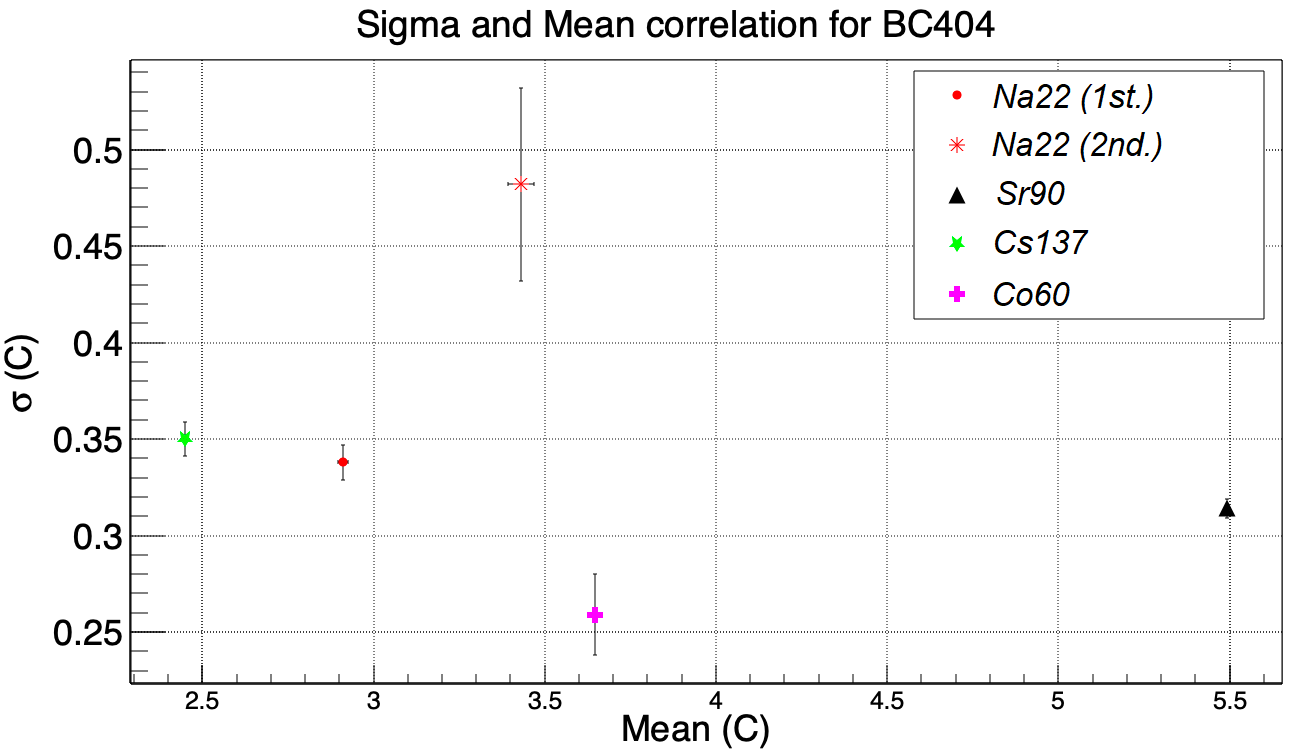}
	\includegraphics[width=0.49\textwidth]{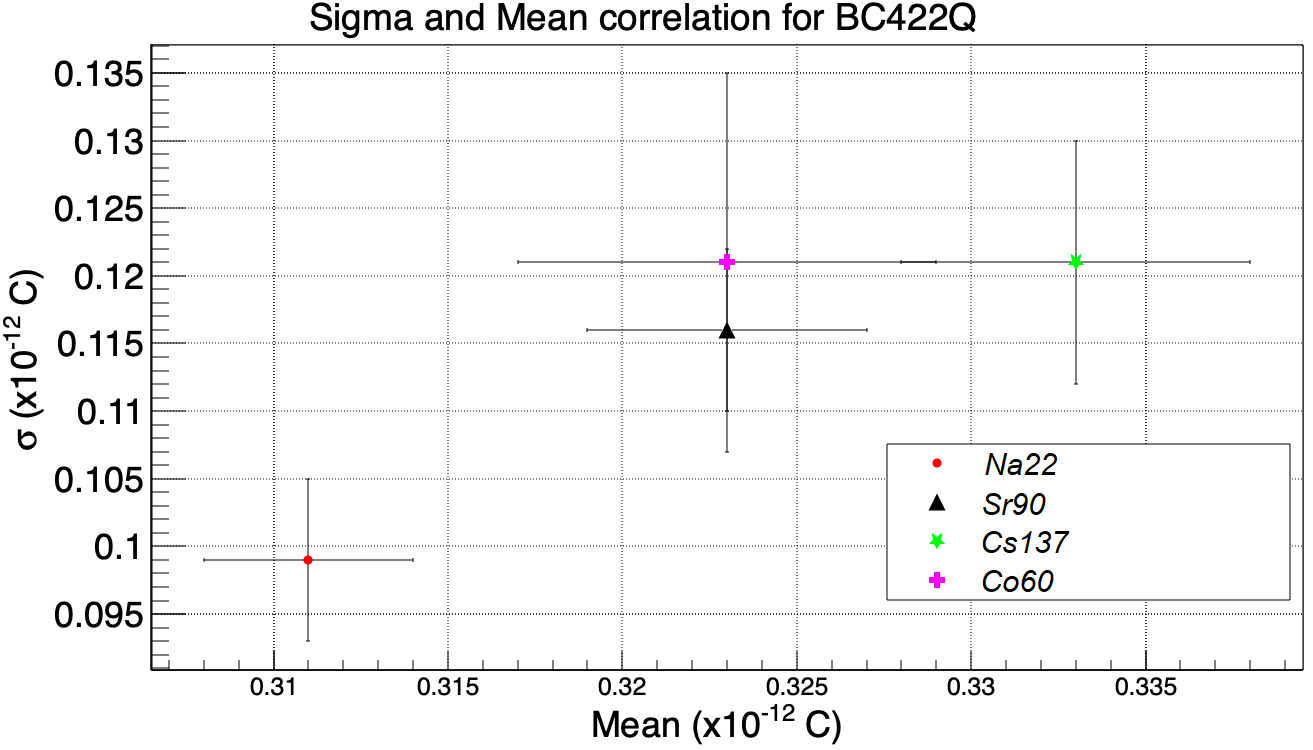}
	\caption{Mean-$\sigma$ relation from the Gaussian Fit for studied radiation sources with BC404 (up) 
	and BC422Q (down) scintillators.}\label{meansigma}
	\end{figure}
\vspace{1cm}
	\section{Conclusions}\label{4}
A relation between the fast and standard output signals of a SensL SiPM photo-sensor was found. For a given pulse, by using the fast output signal, we were able to reconstruct the deposited charge. This value is in good concordance with the estimated charge using the standard pulse. We also observed that the best agreement between the reconstructed charged from the fast output signal and the one  from the standard output signal of a SensL SiPM photo-sensor is obtained for the BC422Q plastic scintillator. This results shows that it may be possible to develop a trigger system based on plastic scintillator material and a SensL SiPM photo-sensor with an excellent time resolution where also the information of electric charge can be reconstructed using the fast output signal of such photo-sensor.
	%Then, it is possible to estimate the deposited charge, instead of the standard output.
%\textcolor{blue}{	From Table \ref{tab:TSKStest} and as the level of significance $\alpha = 0.05$ is always minor than the $p$--values for all cases, the TSKS-test fails to reject the null hypothesis at the $\alpha$ level of $0.05$ so, we conclude that the CDF's and PDF's of the corresponding reconstructed charge deposition are equivalent. }
	
	In average, the conversion factor  among fast and standard charge is
	$0.008\pm0.001\times10^{-12}$ for BC404 and
	$0.012\pm~0.001\times10^{-12}$ for BC422Q. Whence,
	$Q_{S404}/Q_{F404} = 128.205$ and $Q_{S422}/Q_{F422} = 85.106$,  which means that the charge deposition in the BC404 scintillator
	is 1.506 times the one for a BC422Q scintillator. Therefore 
	 (using these thin materials \cite{Lamprou2020}), 
	BC404 scintillator is 1.506  more sensitive  
	than BC422Q scintillator for $^{90}$Sr, $^{60}$Co, $^{137}$Cs and $^{22}$Na radiation
	sources. 
	This result gives the possibility to use the fast pulse from the
	detectors, where time resolution is an important
	restriction \cite{Alvarado2020} and for fast triggering
	systems in Time of Flight (TOF) applications.\\
	The continuity of this work is to estimate the time resolution of both detector configurations. Our plan is to develop a PET, commonly constructed with LYSO crystal, based on plastic scintillator materials with the best possible time resolution which is a key  parameter during the data
	 acquisition chain. For example, the life time of the 
	 isotopes used to acquire brain or heart images is around 2~minutes. Thus, a fast detector response is desired to improve the spatial and time resolution of PET scanners. 
	 
\section*{Acknowledgements}

Support for this work has been received by Consejo Nacional de Ciencia y Tecnolog\'ia grant numbers A1-S-13525 and A1-S-7655. The authors thanks to the  BUAP  Medical Physics  and  Elementary  Particles  Laboratories  for their  kind  hospitality  and  support  during  the  development  of  this  work.

\bibliographystyle{unsrt}  


\end{document}